\documentclass[11pt, preprint]{article}
\usepackage{color} \usepackage{cancel}
\usepackage{float} \usepackage{graphicx}
\usepackage{subfigure} \usepackage{caption}			
\usepackage{dcolumn}
\usepackage[toc,page]{appendix} \usepackage{authblk}	 		
\usepackage{indentfirst}	
\usepackage{authblk}	 		
\usepackage{multirow}     
\usepackage{hyperref}     
\usepackage{amsfonts, amssymb, amsmath}
\usepackage[left=2.5cm,right=2.5cm,top=2.9cm,bottom=2.9cm,a4paper]{geometry}
\usepackage[noadjust]{cite}
\usepackage{filecontents}
\usepackage[T1]{fontenc}
\usepackage[utf8]{inputenc}
\usepackage{palatino,eulervm}

\graphicspath{{./figures/}}
\hypersetup{ colorlinks   = true, 
urlcolor     = blue, 
linkcolor    = blue, 
citecolor    = red 	 
}
\newcommand{\bea}{\begin{eqnarray}} \newcommand{\eea}{\end{eqnarray}}
\newcommand{\beq}{\begin{equation}} \newcommand{\eeq}{\end{equation}}

\title{Observational constraints on 
inflation with the nonlinear sigma fields in light of Plank 2015}
\author[1]{Seoktae Koh\thanks{kundol.koh@jejunu.ac.kr}}
\author[2,3]{Bum-Hoon Lee\thanks{bhl@sogang.ac.kr}}
\author[3]{Gansukh Tumurtushaa\thanks{gansuh@sogang.ac.kr}}
\affil[1]{\it Department of Science Education, Jeju National University, Jeju, 63243, Korea }
\affil[2]{\it Asian Pacific Center for Theoretical Physics, Pohang 37673, Korea}
\affil[3]{\it Center for Quantum Spacetime, Sogang University, Seoul 121-742, Korea \\
Department of Physics, Sogang University, Seoul 121-742, Korea}

\date{\small \today}
\begin{document}

\maketitle

\begin{abstract}
We  study an inflation model with a  nonlinear sigma field
which has $SO(3)$ symmetry. The background solution of
the nonlinear sigma field is proportional to the space
coordinates linearly while keeping the homogeneous
and isotropic background spacetime.
We calculate the observable quantities including the
power spectra of the scalar and  tensor modes,
the spectral indices, the tensor-to-scalar ratio, and the
running of the spectral indices, and then constrain
our model with the Planck 2015 observational data.
\end{abstract}

\section{Introduction}

Since the accidental detection of Cosmic Microwave Background (CMB) radiation by Penzias and Wilson \cite{Penzias:1965wn}, CMB observations by WMAP \cite{Hinshaw:2012aka} and recent Planck satellite \cite{Ade:2013zuv}\cite{Ade:2015xua}
provide the  cosmological data with high precision. Recent Planck data give $n_s \sim 0.968, \,\, r < 0.11$ \cite{Ade:2015xua}
and $\alpha_s =dn_s/d\ln k=-0.003$ \cite{Ade:2015lrj} with 95\% CL.
These cosmological data seem to  favor an inflationary scenario as the solution to the standard big bang problems.
Inflation predicts almost Gaussian and nearly scale invariant spectrum and also provides the seeds for the large scale structure formation
and the primordial gravitational wave. With this coming precision cosmology era, it may be possible to discriminate
inflationary models \cite{Martin:2013nzq} and to obtain the information around Planck time.

The inflation model studied in this paper is based on the nonlinear sigma model which has $SO(3)$ symmetry.
An interesting feature of this model is to take the spatial coordinate dependent background solutions.
Though the space dependent background solutions look like breaking the cosmological principle , {\it i.e.} homogeneity and isotropy,
if the solutions are proportional to the spatial coordinate linearly, the cosmological principle
is preserved. The ansatz of the spatially linear background solution was used for compactifying extra dimension \cite{Omero:1980vx}
and for giving masses to gravitons through of  Higgs mechanism of gravity \cite{'tHooft:2007bf}\cite{Hinterbichler:2011tt}.
The space dependent background solutions are also studied before in Ref. \cite{ArmendarizPicon:2007nr}\cite{Endlich:2012pz} and may be obtained with a two-form gauge field which is dual to a pseudo scalar field \cite{Ohashi:2013mka}.

This model can provide the physical mechanism for the suppression of CMB spectrum at large angular scales \cite{Bennett:2003bz}\cite{Contaldi:2003zv}. In addition, if the nonlinear sigma field is not coupled to an inflaton, the comoving horizon stays constant in the early phase of the inflation because of the space-dependent solution term. This implies that there exists the  minimum  $k$ mode, $k_{min}$. Because the modes satisfying $k < k_{min}$ do not contribute to the power spectrum, this loss of $k$ modes results in the lack of the  power spectrum of the tensor  \cite{koh2015} as well as the scalar modes \cite{Kouwn:2014aia}.
This is a generic feature of this model. In light of recent Planck data \cite{Ade:2015lrj}, we constrain on the model parameters for this inflation model.

This paper is organized as follows: In section \ref{model}, we briefly introduce the  inflation model with the nonlinear sigma fields.
In section \ref{phin}, the background solutions for $V \sim \phi^n$ are provided through the iteration method
assuming $\xi$ is small enough. With these solutions, $e$-folding number is obtained. The power spectra
for the scalar and tensor modes are presented in section \ref{perturbation} and then we can calculate
the observation variables in section \ref{observation}. Also, the observation variables are constrained in light
of Planck 2015 data in section \ref{observation}. Finally, we summarize our results in section \ref{conclusion}.

\section{Background Evolution} \label{model}

In this section we describe the background evolution of the  inflation model
with the nonlinear sigma fields ~\cite{Koh:2013msa, Kouwn:2014aia}.
The dynamics of a scalar field, $\phi$, with an additional triad of scalar fields, $\sigma^a$, is governed by the action
\bea
\mathcal{S} = \int d^4x \sqrt{-g} \biggl[\frac{1}{2\kappa^2}R -\frac{1}{2} g^{\mu\nu} \partial_{\mu}\phi \partial_{\nu}\phi - V(\phi) -\frac{1}{2}g^{\mu\nu}\delta_{ab} \partial_{\mu}\sigma^a \partial_{\nu}\sigma^b  \biggr]\,, \label{eq:act}
\eea
where $\sigma^a$'s have an $SO(3)$ symmetry,  the indices $a$ and $b$ run from 1 to 3  and $\kappa^2 = 8\pi G$. Varying the action Eq. (\ref{eq:act}) with respect to the metric $g_{\mu\nu}$
yields the Einstein equation
\begin{align}
R_{\mu\nu} -\frac{1}{2}g_{\mu\nu} R =\frac{\kappa^2}{3} T_{\mu\nu}\,,
\end{align}
where the energy-momentum tensor is given by
\begin{align} \label{emtensor}
T_{\mu\nu} =& \partial_\mu \phi \partial_\nu \phi
+ \partial_\mu \sigma^a \partial_\nu \sigma^a
-g_{\mu\nu}\Biggr[{1\over2} g^{\alpha\beta}\partial_\alpha\sigma^a \partial_\beta\sigma^a
+{1\over2} \, g^{\alpha\beta}\partial_\alpha \phi \partial_\beta \phi  +V(\phi)
\Biggr],
\end{align}
and with respect to the scalar fields $\phi$ and $\sigma^a$ yields the equations of motion for both $\phi$ and $\sigma^a$
\begin{align}
&\partial_\mu\partial^\mu \phi -V_\phi=0,
\\
&{1\over\sqrt{-g}} \partial_\mu \Big( \sqrt{-g} g^{\mu\nu}\partial_\nu \sigma^a \Big)=0\,, \label{eom_sigma}
\end{align}
where $V_{\phi}\equiv dV/d\phi$. Assuming the flat Friedmann-Lemaitre-Robertson-Walker (FLRW) metric for $g_{\mu\nu}$,
\begin{align}
ds^2 = -dt^2 + a^2 \delta_{ij} dx^i dx^j,
\end{align}
where $a(t)$ is a scale factor, and requiring the cosmological principle of homogeneity and isotropy to be preserved, we choose a spatially linear background solution for $\sigma^a$~\cite{ArmendarizPicon:2007nr, Endlich:2012pz, 'tHooft:2007bf, Lee:2009zv, Koh:2013msa, Kouwn:2014aia}
\begin{align}\label{eq:sigma}
\sigma^a = \xi x^i \delta_i^a,
\end{align}
where $\xi$ is an arbitrary constant parameter of $[M_p^2]$ dimension, which satisfies the equations of motion  Eq. (\ref{eom_sigma}) \footnote{We use $\xi$ instead of $\alpha$
which was used in Ref.~\cite{Koh:2013msa} to avoid confusion with the running spectral indices, $\alpha$, that we will use later.}.
The background dynamics of this system  yields the following equations of motion for the homogeneous scalar field, $\phi$, and the scale factor $a(t)$
\begin{align}
&  H^2 = \frac{\kappa^2}{3}\biggl(\frac{1}{2}\dot{\phi}^2+V+\frac{3}{2a^2}\xi^2\biggr),\label{bgeq1} \\
& \dot{H} = -\frac{\kappa^2}{2} \biggl(\dot{\phi}^2+\frac{\xi^2}{a^2}\biggr),\label{bgeq2} \\
& \ddot{\phi}+3H\dot{\phi}+V_{\phi} = 0, \label{bgeq3}
\end{align}
where $H\equiv\dot{a}/a$.
The energy density and pressure  from  (\ref{emtensor}) with  (\ref{eq:sigma})
are given by
\begin{align}\label{rho}
\rho =& \frac{1}{2}\dot{\phi}_0^2 + V + \rho_{\sigma}, \\
p =& \frac{1}{2} \dot{\phi}_0^2 - V + p_{\sigma}\label{press},
\end{align}
where
\begin{align}
\rho_{\sigma}=&\frac{3\xi^2}{2a^2} \quad \text{and} \quad p_{\sigma}= - \frac{\xi^2}{2a^2}.
\label{rhosigma}
\end{align}

In the early phase of the evolution of the Universe, $\xi$ dependent term can dominate over the kinetic and the potential terms due to the factor of $1/a^2$. Then the resulting equation of state parameter in those $\xi$ dominant phase  from Eq. (\ref{rhosigma}),
$w_{\sigma}\equiv {p_{\sigma}}/{\rho_{\sigma}} = -\frac{1}{3}$ , shows that inflation does not occur in the early phase of the evolution of the Universe. As the Universe evolves, $\xi$ dependent term decays away very quickly and then the potential term
of the scalar field $\phi$ begins to dominate such that the spacetime experiences an accelerated expansion.

In this work, in order to have an inflationary solution, we assume that $\xi$ parameter is small enough, hence we can neglect the pre-inflation phase. Therefore, during the phase of our interests, the initial state for the quantum fluctuations stays in the Bunch-Davies vacuum state and the potential term dominates over $\xi$ dependent term such that usual slow-roll inflation takes place. Although, there still remains small contribution from the $\xi$ dependent term~\cite{Koh:2013msa}.

\section{Slow-Roll Inflation model with $V \sim \phi^n$ potential} \label{phin}

As we learned in previous section that the spacetime experiences an accelerated expansion, $\ddot{a}>0$, if and only if the $\xi$ is small enough such that the potential energy of scalar field dominates over its kinetic energy,
\begin{align}\label{eq:sl_co1}
\dot{\phi}^2\ll V.
\end{align}
The accelerated expansion can be sustained for a sufficiently long period of time if the second order time derivatives of $\phi$ is small enough
\begin{align}\label{eq:sl_co2}
|\ddot{\phi}|\ll|3H\dot{\phi}|,\,|V_{\phi}|.
\end{align}
Therefore, in this section, we consider the slow-roll inflation model with the power law potential
\begin{align}
V(\phi) = \lambda M_p^4 \left(\frac{\phi}{M_p}\right)^n,
\label{potential}
\end{align}
where $\lambda$ is an arbitrary dimensionless parameter. The background equations of motion,  (\ref{bgeq1}) and  (\ref{bgeq3}), with the slow-roll approximations, Eqs. (\ref{eq:sl_co1})--(\ref{eq:sl_co2}), and with the inflaton potential Eq. (\ref{potential}) yield
\begin{align}
H^2 \simeq& \frac{\kappa^2}{3} \left( \lambda M_p^4 \left(\frac{\phi}{M_p}\right)^n
+\frac{3{\xi}^2}{2a^2} \right),
\label{bgeq_sr1}
\\
3H\dot{\phi} +&  n\lambda M_p^3 \left(\frac{\phi}{M_p}\right)^{n-1} \simeq 0.
\end{align}
We define the slow-roll parameters following \cite{Koh:2013msa} to reflect the slow-roll conditions, in Eqs. (\ref{eq:sl_co1})--(\ref{eq:sl_co2}), as
\begin{align}\label{eq:sl_param}
\epsilon\equiv\frac{\kappa^2 \dot{\phi}^2}{H^2}\,, \qquad \eta \equiv \frac{V_{\phi\phi}}{3H^2},
\end{align}
and inflation requires smallness of the slow-roll parameters. Inflation ends when the slow-roll conditions are violated $\epsilon(\phi_{e})=1$ where $\phi_{e}$ is the field value at the end of inflation. We compute the $e$-folding number, following \cite{Koh:2013msa}, before the end of inflation
\begin{align}
\mathcal{N} =& \kappa^2\int^{\phi_i}_{\phi_e}
\left( \frac{V}{V_{\phi}} + \frac{3^2}{2a^2 V_{\phi}} \right) d\phi
\nonumber \\
=& \frac{\kappa^2}{n}  \int^{\phi_i}_{\phi_e} \phi d\phi
+ \frac{3\kappa^2 {\xi}^2 M_p^{n-4}}{2n\lambda} \int^{\phi_i}_{\phi_e}
\frac{1}{a^2 \phi^{n-1}} d\phi.
\label{efold}
\end{align}
Since $\xi$ is small which is consistent with observations \cite{Koh:2013msa}, we can find the background analytic solutions through the iteration method. We can expand $a$ and $\phi$ up to ${\xi}^2$  order
\begin{align}
a(t) =&\,\, a_0(t) + \xi^2 a_1(t) + \mathcal{O}(\xi^4), \\
\phi(t) =& \,\, \phi_0 (t) + \xi^2 \phi_1 (t) +\mathcal{O} (\xi^4).\label{eq:phi}
\end{align}
 Because the second  term in (\ref{efold}) is already order of $\xi^2$,
  it is enough to consider only $0$th order solutions for both $\phi(t)$ and $a(t)$.
  At zeroth order, we obtain  for general $n$ $(n\neq 4)$
\begin{align}\label{eq:phi0}
\phi_0 (t) =& 2^{\frac{2}{n-4}} \left\{ (n-4) (\beta t - c_1) \right\}^{-\frac{2}{n-4}},
\\
a_0(t) =& a_i e^{-\frac{\kappa^2}{2n} (\phi_{0}^2-\phi_i^2)}, \label{eq:a0}
\end{align}
and for $n=4$
\begin{align}
\phi_0 (t) = \phi_i e^{-\beta t},
\end{align}
where $\beta = \sqrt{n^2 \lambda M_p^{4-n}/(3\kappa^2)}$.

Then the $e$-folding number Eq. (\ref{efold}) up to $\xi^2$ order becomes
\begin{align}\label{eq:efold}
&&\mathcal{N}
= \frac{\kappa^2}{2n} (\phi^2 -\phi_e^2)
+  \frac{ {\xi}^2 n}{2\alpha^2 } \frac{1}{a_i^2 e^{\frac{\kappa^2}{n}\phi_i^2}}
\left\{-\frac{1}{2}\phi_0^{2-n} \left(-\frac{\kappa^2 \phi_0^2}{n} \right)^{n/2-1}
\Gamma \left(1-\frac{n}{2},-\frac{\kappa^2 \phi_0^2}{n} \right) \right\}\,,
\end{align}
where $\Gamma(a,z)$ is the incomplete Gamma function.
If we take $\phi_0 \sim 15 M_p$, then
because $\kappa^2 \phi_0^2/n \sim 8\pi (15 M_p)^2/ n M_p^2 \sim 10^4/2n$,
we can approximate $x \equiv -\kappa^2 \phi_0^2/n \rightarrow \infty$. Using
the asymptotic property of the incomplete Gamma function as $x \rightarrow \infty$
\begin{align}
\Gamma(s,x) \rightarrow& x^{s-1} e^{-x} \,\,\, \text{as} \,\,\, x\rightarrow \infty\,,\nonumber
\end{align}
we use asymptotically
\begin{align}\label{eq:gammafunc}
\Gamma \left(1-\frac{n}{2},-\frac{\kappa^2}{n} \phi_0^2 \right)
\simeq \left(-\frac{\kappa^2 \phi_0^2}{n}
\right)^{-\frac{n}{2}} e^{\frac{\kappa^2}{n} \phi_0^2}.
\end{align}
After substituting Eq. (\ref{eq:phi}) and Eq. (\ref{eq:gammafunc}) into Eq. (\ref{eq:efold}), we obtain
\begin{align}\label{eq:efolding}
\mathcal{N}\simeq &\frac{\kappa^2}{2n}\biggl[\phi_0^2+\frac{2{\xi}^2}{M_p^4} \left(M_p^4 \phi_0\phi_1 +\frac{3n}{4\kappa^2 \lambda M_p^{-n}}\frac{1}{a_0^2 \phi_0^n} \right) \biggr],
\end{align}
where we neglected $\phi_e^2$ because we assume that the value of the scalar field at the end of inflation is much smaller than that of the beginning, $\phi_e\ll\phi$.

For simplicity, we consider only $n=2$ case in this work. Next leading order solution for $\phi_1(t)$ can be obtained as
\begin{align}
\phi_1(t) =& \frac{3M_p^2}{4\lambda M_p^4}\frac{1}{a_0^2\phi_0}\left[1-\sqrt{2\kappa^2}\phi_0 F\left(\sqrt{\frac{\kappa^2}{2}}\phi_0\right)\right]\simeq 0,
\label{approx_n2}
\end{align}
where $F(x)$ is a Dawson function and the last $"\simeq "$ holds for $\sqrt{\kappa^2/2}\phi_0\rightarrow\infty$ because $F'(x)=1-2xF(x)\rightarrow0$ as $x\rightarrow\infty$.

Therefore, the $e$-folding number, Eq. (\ref{eq:efolding}), becomes
\begin{align}
\mathcal{N}\approx&\frac{\kappa^2}{4}\biggl(\phi_0^2+\frac{M_p^4 \delta}{a_0^2\phi_0^2}\biggr), \label{efold_n2}
\end{align}
where we introduce a dimensionless variable
 $\delta \equiv \frac{3{\xi}^2}{2\lambda \kappa^2 M_p^6}$ and which is much smaller than unity.
Because $\delta\ll1$, we expand $\varphi_N \equiv \phi_0^2$ as a function of
$\mathcal{N}$ as
\begin{align}\label{eq:vafN}
\varphi_{\mathcal{N}} =& \varphi^0 (\mathcal{N}) + \delta \varphi^1 (\mathcal{N}) +\mathcal{O}(\delta^2).
\end{align}
Then Eq. (\ref{efold_n2}) yields up to linear order in $\delta$
\begin{align}\label{eq:Nofvaf}
\mathcal{N} =& \frac{\kappa^2}{4} \biggl[ \varphi^0 (\mathcal{N})
+\delta \varphi^1(\mathcal{N}) + \delta \frac{1}{a_i^2 e^{\frac{\kappa^2}{4}\phi_i^2}
e^{-\frac{\kappa^2}{4}\varphi^0} \varphi^0} \biggr].
\end{align}
We obtain the following expressions for the zeroth order in $\delta$,
\begin{align}\label{eq:vaf0}
\mathcal{N} =  \frac{\kappa^2}{4} \varphi^0 (\mathcal{N}),
\end{align}
and for the first order in $\delta$,
\begin{align}\label{eq:vaf1}
0 =& \varphi^1 (\mathcal{N}) +\frac{1}{a_i^2 e^{\frac{\kappa^2}{4}\phi_i^2}e^{-\frac{\kappa^2}{4}\varphi^0} \varphi^0}.
\end{align}
By substituting Eqs. (\ref{eq:vaf0})--(\ref{eq:vaf1}) into Eq. (\ref{eq:vafN}), we obtain the scalar field as function of $\mathcal{N}$, the number of $e$-folds.
\begin{align}
\phi_0^2=&\varphi_{\mathcal{N}}=\frac{4}{\kappa^2}\mathcal{N}-\frac{\kappa^2}{4}\frac{\delta}{a_i^2e^{\frac{\kappa^2}{4}\phi_i^2}e^{-\mathcal{N}}\mathcal{N}}.
\end{align}

We reexpress the slow-roll parameters, Eq. (\ref{eq:sl_param}),
in terms of the inflaton potential for $n=2$ in Eq. (\ref{potential})
\begin{align}
\epsilon =&  \frac{V_{\phi}^2}{2\kappa^2 V^2} \left( 1- \frac{3{\xi}^2}{a^2 V} \right)
=\frac{2}{\kappa^2 \phi^2} \left(1- \frac{3{\xi}^2}{\lambda M_p^2 a^2 \phi^2} \right),
\label{srp1}
\\
\eta =&  \frac{V_{\phi\phi}}{\kappa^2 V}\left( 1- \frac{3{\xi}^2}{2a^2 V} \right)
= \frac{2}{\kappa^2 \phi^2} \left(1- \frac{3{\xi}^2}{2\lambda M_p^2 a^2 \phi^2}\right).
\label{srp2}
\end{align}
If we use the same approximation for $n=2$  as in Eq. (\ref{approx_n2}), we
can replace $\phi$ as $\phi_0$ which is zeroth order in $\xi^2$
\begin{align}\label{eq:sl_eps}
\epsilon =& \frac{2}{\kappa^2 \phi_0^2} \left(1- \frac{3{\xi}^2}{
\lambda M_p^2 a_i^2 e^{\frac{\kappa^2}{4}\phi_i^2}
 e^{-\frac{\kappa^2}{4}\phi_0^2} \phi_0^2} \right),
\\
\eta =& \frac{2}{\kappa^2 \phi_0^2} \left(1- \frac{3{\xi}^2}{2\lambda M_p^2
a_i^2 e^{\frac{\kappa^2}{4}\phi_i^2}
 e^{-\frac{\kappa^2}{4}\phi_0^2} \phi_0^2}
\right).\label{eq:sl_eta}
\end{align}
In terms of $\mathcal{N}$, Eqs. (\ref{eq:sl_eps})--(\ref{eq:sl_eta}) can be written as
\begin{align}\label{eq:sl_param1}
\epsilon=\frac{1}{2 \mathcal{N}}\left(1-\frac{3 \kappa ^2 \xi ^2 (8 \mathcal{N}-1) }{32\lambda M_p^2 a_i^2 e^{\frac{\kappa ^2 \phi_i^2}{4}} e^{\mathcal{-N}} \mathcal{N}^2}\right),\\
\eta=\frac{1}{2 \mathcal{N}}\left(1-\frac{3 \kappa ^2 \xi ^2 (4 \mathcal{N}-1) }{32\lambda M_p^2 a_i^2 e^{\frac{\kappa ^2 \phi_i^2}{4}} e^{\mathcal{-N}} \mathcal{N}^2}\right).\label{eq:sl_param2}
\end{align}

\section{The linearized equations of motion}  \label{perturbation}

We will briefly review the linear perturbations in this model  (see the details in
\cite{Koh:2013msa} with $f(\phi)=1$).
 With the perturbed metric in a conformal Newtonian gauge
\begin{align}
ds^2 = - a^2(\tau) [(1+2A) d\tau^2 + \{ (1-2\psi) \delta_{ij} +h_{ij} \}dx^i dx^j],
\end{align}
where ${h^i}_{j;i} = 0 = {h^i}_i$,
and
 decomposing $\phi(t,{\bf x}) = \phi(t) + \delta \phi(t,{\bf x}),\,\,
 \sigma^a = \xi  x^a + \partial^a u$, we can construct  two gauge invariant
 variables for the scalar modes which can be represented as:
 \begin{align}
 Q_{\phi} =& \,\,\delta \phi - \psi, \quad Q_u = \,\,u .
 \end{align}
Because we are interested in constraining on model parameters from
the Planck data using $n_s, \,\,r$ and $\alpha_s$, we will put the
isocurvature mode aside in this work.

After Fourier transforming of $Q_{\phi}$ and $h_{ij}$  to the momentum space
\begin{align}
Q_{\phi}(\tau) =&  \frac{1}{a} \int d^3k u_k (\tau) e^{i{\bf k}\cdot {\bf x}}, \\
h_{ij} (\tau) =& 2\frac{\sqrt{8\pi G}}{a}
\int d^3k v_k(\tau) e^{i{\bf k}\cdot {\bf x}},
\end{align}
we can obtain the Sasaki-Mukhanov equation for the scalar and tensor modes
\begin{align}
u_k'' \,+& \left( k_s^2-\frac{\mu_s^2 -\frac{1}{4}}{\tau^2} \right) u_k = 0,
\\
v_k'' \,+& \left(k_t^2 - \frac{\mu_t^2-\frac{1}{4}}{\tau^2} \right) v_k = 0,
\end{align}
where\footnote{We
have corrected the numerical factor in $k_t^2$ in Ref. \cite{Koh:2013msa}.}
\begin{align}
k_s^2 =& k^2 - \frac{{\xi}^2}{6 M_p^2}, \quad
k_t^2 =  k^2 + \frac{11{\xi}^2}{6 M_p^2}
\label{eff_ks}
\\
\mu_s =& \frac{3}{2} +3\epsilon - \eta, \quad
\mu_t = \frac{3}{2} + \epsilon
\end{align}

Choosing the Bunch-Davies  vacuum for an initial state
at $\tau \rightarrow -\infty$  by taking the positive frequency modes,
we obtain the exact solutions for the scalar and tensor modes
\begin{align}
u_k =& \frac{\sqrt{\pi}}{2} e^{i\frac{\pi}{2}(\mu_s+\frac{1}{2})}\sqrt{-\tau}
H^{(1)}_{\mu_s} (-k_s \tau),
\label{uk}
\\
v_k =& \frac{\sqrt{\pi}}{2} e^{i\frac{\pi}{2}(\mu_t+\frac{1}{2})}\sqrt{-\tau}
H^{(1)}_{\mu_t} (-k_t \tau).
\label{vk}
\end{align}
$k_s$ should be real to have a well-defined quantum state, so $k_s^2$ also
should be positive. Then $k^2$ can be constrained from the expression
of $k_s^2$ in (\ref{eff_ks})
\begin{align}
k_s^2 \geq \frac{{\xi}^2}{6 M_p^2}.
\end{align}
The existence of the cutoff scale, $k_{min} \equiv {\xi}/\sqrt{6}M_p$,
implies the lack of the power  in  the power spectrum \cite{Kouwn:2014aia}
and results in the suppression of the angular power spectrum in
CMB. The cut-off scale for the scalar modes is shown explicitly in (\ref{eff_ks}).
There exists, however,  another cutoff scale  in the power spectrum for the
scalar modes and that for the tensor modes as well
which is not shown explicitly \cite{koh2015}. This cutoff originates
from the fact that  the comoving horizon is given by $aH \sim \kappa \xi/\sqrt{2} $
in (\ref{bgeq_sr1}) in the early phase when the $\xi$ term is dominant.
Therefore the model satisfying $k < \kappa \xi/\sqrt{2} $ cannot contribute to
the observed power spectra for the scalar and tensor modes.
This cutoff scale seems to be a very generic feature of this inflation scenario
which is motivated from the nonlinear sigma model.

$u_k$ and $v_k$ in (\ref{uk}) and (\ref{vk}) become in
the large scale limit $x = k\tau \ll 1$
\begin{align}
u_k \simeq&  e^{i\frac{\pi}{2}(\mu_s+\frac{1}{2})} 2^{\mu_s-\frac{3}{2}}
 \frac{\Gamma(\mu_s)}{ i\Gamma(\frac{3}{2})}
\frac{1}{\sqrt{2k_s}}
(-k_s\tau)^{-\mu_s +\frac{1}{2}},
\\
v_k \simeq&  e^{i\frac{\pi}{2}(\mu_t+\frac{1}{2})} 2^{\mu_t-\frac{3}{2}}
 \frac{\Gamma(\mu_t)}{ i\Gamma(\frac{3}{2})}
\frac{1}{\sqrt{2k_t}}
(-k_t\tau)^{-\mu_t +\frac{1}{2}},
\end{align}
where we have used
\begin{align}
H^{(1)}_{\mu} \sim \frac{-2}{1-e^{2i\mu\pi}}
\frac{e^{i\mu\pi}}{\Gamma(1-\mu)} \left(\frac{x}{2} \right)^{-\mu}.
\end{align}

Since $\sigma^a \propto \mathcal{O}(\xi)$,
we can expect $\delta \sigma^a  = \partial^i u = \partial^i Q_u
\propto \mathcal{O}(\xi^2)$.  If we keep only $\xi^2$ order, curvature perturbations become\footnote{
Here we recover $8\pi G = \frac{8\pi}{M_p^2}$.
In \cite{Koh:2013msa}, because they used the reduced Planck unit,
we multiply $\sqrt{8\pi}$ in (\ref{rnq})}
\begin{align}
\mathcal{R} \simeq  \frac{\sqrt{4\pi}}{\sqrt{\epsilon} M_p } \left( 1
-\frac{{\xi}^2}{2\epsilon  M_p^2 a^2 H^2} \right) Q_{\phi}.
\label{rnq}
\end{align}

The power spectra in the large scale limit of the curvature
perturbation $\mathcal{R}$ and the tensor modes yield
\begin{align}
\mathcal{P}_{\mathcal{R}}(k) =& \frac{k^3}{2\pi^2} |\mathcal{R}_k|^2
= \frac{2k^3}{\pi\epsilon M_p^2 a^2} \left(1-\frac{{\xi}^2}{\epsilon M_p^2 a^2 H^2}
\right) |u_k|^2
\nonumber \\
\simeq& \frac{H^2}{\pi \epsilon M_p^2}
 \left(1 +2C\delta
 -2 \epsilon  -\frac{{\xi}^2}{\epsilon M_p^2 a^2 H^2}
 \right)  \left(\frac{k}{aH} \right)^{-6\epsilon + 2\eta},
\label{powers}
\\
\mathcal{P}_{T} (k) =& 2\mathcal{P}_h
= \frac{k^3}{2\pi^2} \frac{64\pi}{M_p^2 a^2} |v_k|^2
\nonumber \\
\simeq& \frac{16 H^2}{\pi M_p^2}
\left(1 + (2C-2) \epsilon - \frac{33{\xi}^2}{12 M_p^2 k^2}
-\frac{{\xi}^2}{3M_p^2 a^2 H^2} \right)
\left(\frac{k}{aH}\right)^{-2\epsilon},
\label{powert}
\end{align}
where the numerical factor ``2'' in (\ref{powert})  comes from the two polarization states and
$\delta = 3\epsilon -\eta, \,\, C = 2-\gamma - \ln 2$
and $\gamma \approx 0.5772$ is a Euler-Mascheroni constant.
We have also used
\begin{align}
\tau =-\frac{1}{aH} \left( 1+\epsilon +\frac{{\xi}^2}{6M_p^2 a^2
H^2} \right).
\end{align}
Note that (\ref{powers}) is valid only in the range of $\xi^2$ \cite{Koh:2013msa}
\begin{align}
{\xi}^2 \leq \epsilon k^2 M_p^2.
\label{const}
\end{align}

\section{Observational constraint in light of Plank 2015} \label{observation}

 We only consider  $n=2$ potential in (\ref{potential})
for an observational constraint on our model.
From (\ref{powers}) and (\ref{powert}), we compute the spectral indices, the tensor-to-scalar ratio and the running spectral indices at the horizon crossing time with the help of slow-roll parameters in Eqs. (\ref{eq:sl_param1})--(\ref{eq:sl_param2}).
\begin{align}
n_s-1 =& \frac{d\ln \mathcal{P}_{\mathcal{R}}(k)}{d\ln k} \biggl|_{k=aH}
= 2\eta -6\epsilon + \frac{2{\xi}^2}{M_p^2 \epsilon k^2}\label{eq:ns}\, ,
\\
\alpha_s =& \frac{d n_s}{d\ln k} \biggl|_{k=aH}=
\frac{-4{\xi}^2}{M_p^2 \epsilon k^2}\label{eq:alphas} \, ,
\\
n_t =& \frac{d\ln \mathcal{P}_T (k)}{d\ln k} \biggl|_{k=aH}
= -2\epsilon + \frac{31{\xi}^2}{6M_p^2 k^2} \, ,
\\
\alpha_t =& \frac{d n_t}{d\ln k}\biggl|_{k=aH} = -\frac{31{\xi}^2}{3M_p^2 k^2} \, ,
\\
r \equiv & \frac{\mathcal{P}_T}{\mathcal{P}_{\mathcal{R}}}\biggl|_{k=aH}
= 16\epsilon \left( 1 - (4C -2)\epsilon +2C\eta
+\frac{{\xi}^2}{M_p^2 \epsilon k^2} \right)\label{eq:ttsrr}\, .
\end{align}
Because ${\xi}^2/M_p^2 \epsilon^2 k^2 > 0$,
$\alpha_s$ and $\alpha_t$ give  the negative value
and if we take into account of (\ref{const}),
they are order of $\epsilon$ and $\epsilon^2$, respectively.

\begin{figure}[H]
   \centering
   \subfigure[$k=10^{-1}$]
    {\includegraphics[scale=0.8]{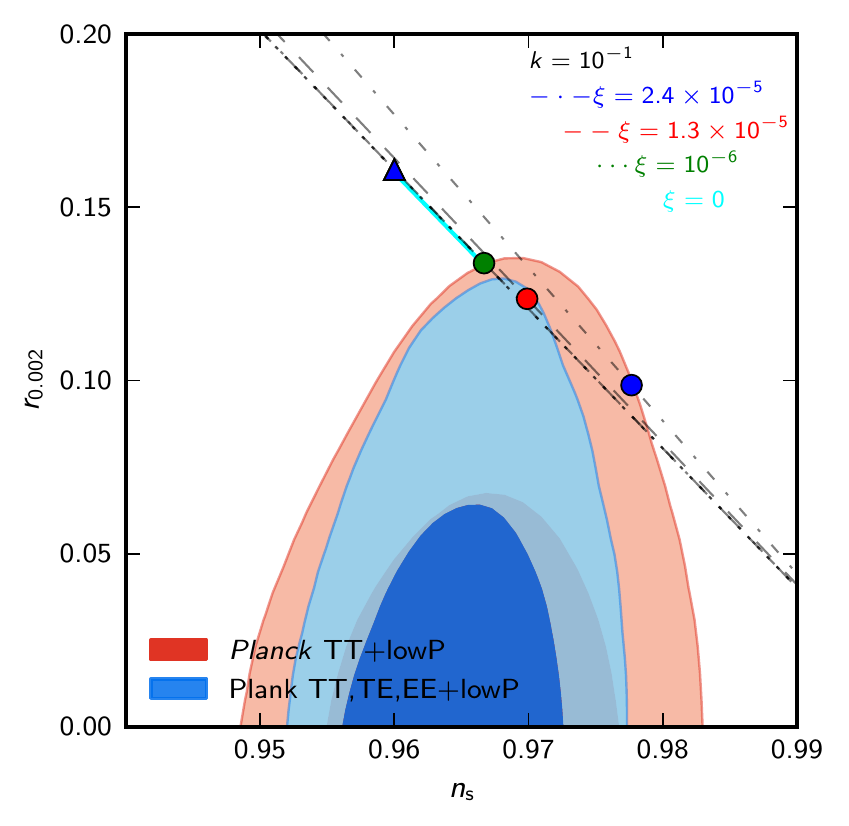}\label{sl_ke-1}}
  \subfigure[$k=10^{-2}$]
    {\includegraphics[scale=0.8]{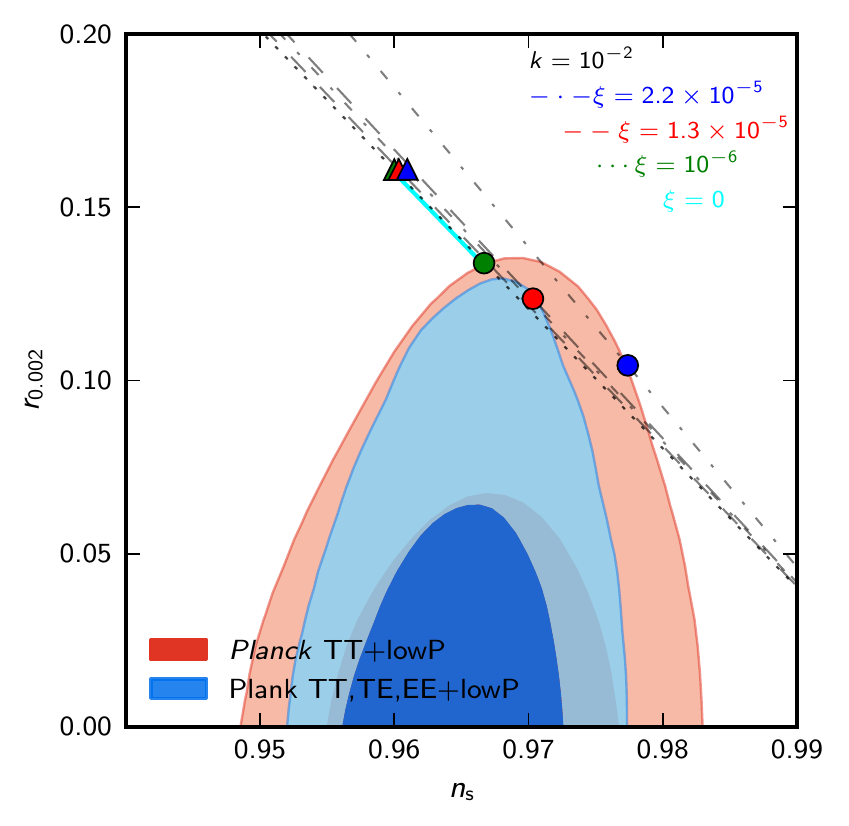}\label{sl_ke-2}}
    \\
    \subfigure[$k=10^{-3}$]
    {\includegraphics[scale=0.8]{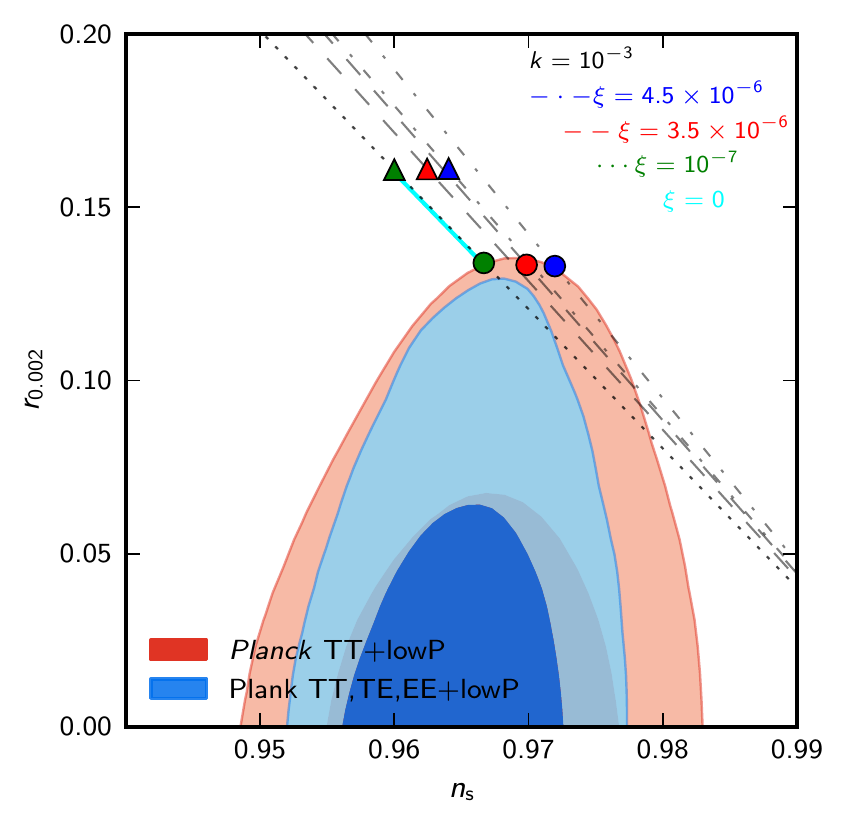}\label{sl_ke-3}}
    \subfigure[$k=10^{-4}$]
    {\includegraphics[scale=0.8]{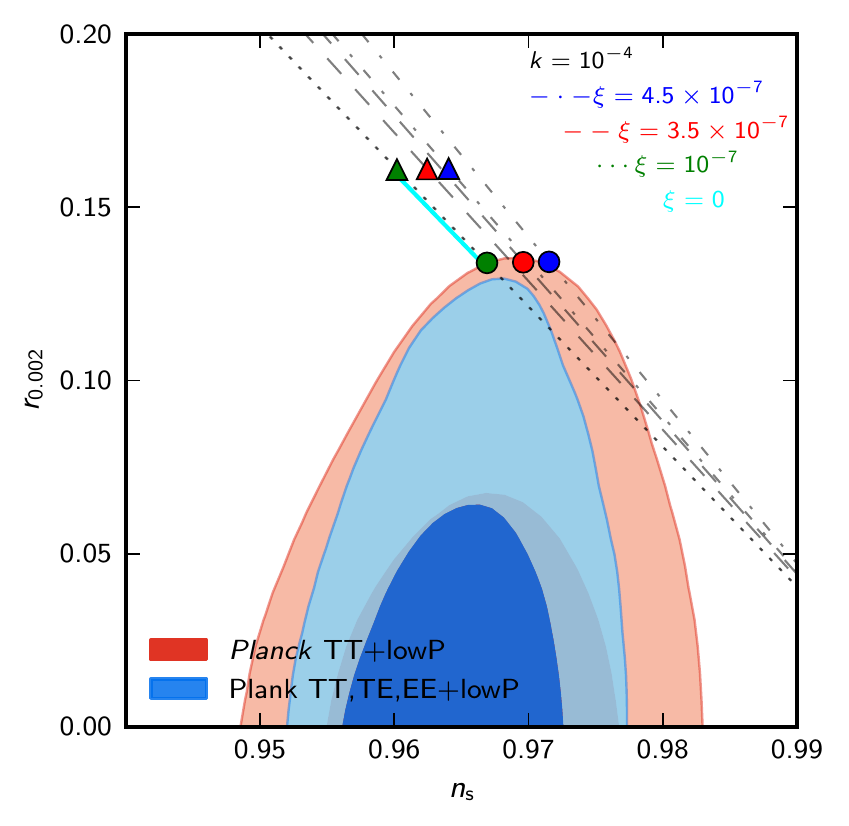}\label{sl_ke-4}}
    \caption{We plot $n_s$ vs. $r$ for $\lambda=0.5\times10^{-12}$, $\phi_i=16M_p$, $a_i=1$ and $M_p^2=1$. The circle represents $N=60$ while the triangle corresponds to $N=50$.}
   \label{fig1}
\end{figure}

For the numerical results below, we set $\lambda=0.5\times10^{-12}$, $\phi_i=16M_p$, $a_i=1$ and $M_p^2=1$.
We plot $n_s$ vs. $r$ in Fig. \ref{fig1} where, then throughout this paper, the circle represents $N=60$ while the triangle corresponds to $N=50$.
The usual single field inflation predictions  correspond to $\xi = 0$
and the  model expectation for $N=50$ situates outside of $2\sigma$ contour while it situates at the edge of $2\sigma$ contour for $N=60$ \cite{Ade:2015lrj}. If  we turn on $\xi$ value, our model shows different results depending on $\xi$ as well as $k$ value. The $k$ value decreases from $10^{-1}$ to $10^{-4}$ from Fig.~\ref{sl_ke-1} to Fig.~\ref{sl_ke-4}, respectively. In Fig.\ref{sl_ke-1} where we set $k=10^{-1}$, the model expectation situates inside $2\sigma$ contour and $\xi$ takes values in interval $10^{-6}<\xi<2.4\times10^{-5}$ for $N=60$. Similarly in Fig.~\ref{sl_ke-2} where we set $k=10^{-2}$, the parameter range of $\xi$ which situates inside $2\sigma$ contour is $10^{-6}<\xi<2.2\times10^{-5}$ for $N=60$. In Fig.~\ref{sl_ke-3} where we set $k=10^{-3}$, the parameter range of $\xi$ that keeps the model expectation inside $2\sigma$ contour yields $10^{-7}<\xi<4.5\times10^{-6}$ for $N=60$; and $\xi$ takes values in interval $10^{-7}<\xi<4.5\times10^{-7}$ for $N=60$ in Fig.~\ref{sl_ke-4} where we set $k=10^{-4}$.
From Fig.~\ref{fig1}, we see that the model expectation for $N=60$ shows better consistency with the data rather than that of $N=50$ for which our results situate outside of $2\sigma$ contour. When model parameter satisfies condition $\xi\leq10^{-6}$ for both $k=10^{-1}$ and $k=10^{-2}$; and $\xi\leq10^{-7}$ for both $k=10^{-3}$ and $k=10^{-4}$, we produce same result as $\xi=0$ case. As is seen in Fig.~\ref{fig1}, we obtain parameter ranges of $\xi$ for different $k$ such that the model expectation is consistent with the observational data~\cite{Ade:2015lrj} up to $95\%$ confidence level.

\begin{figure}[H]
    \centering
    \subfigure[caption]
    {\includegraphics[scale=0.7]{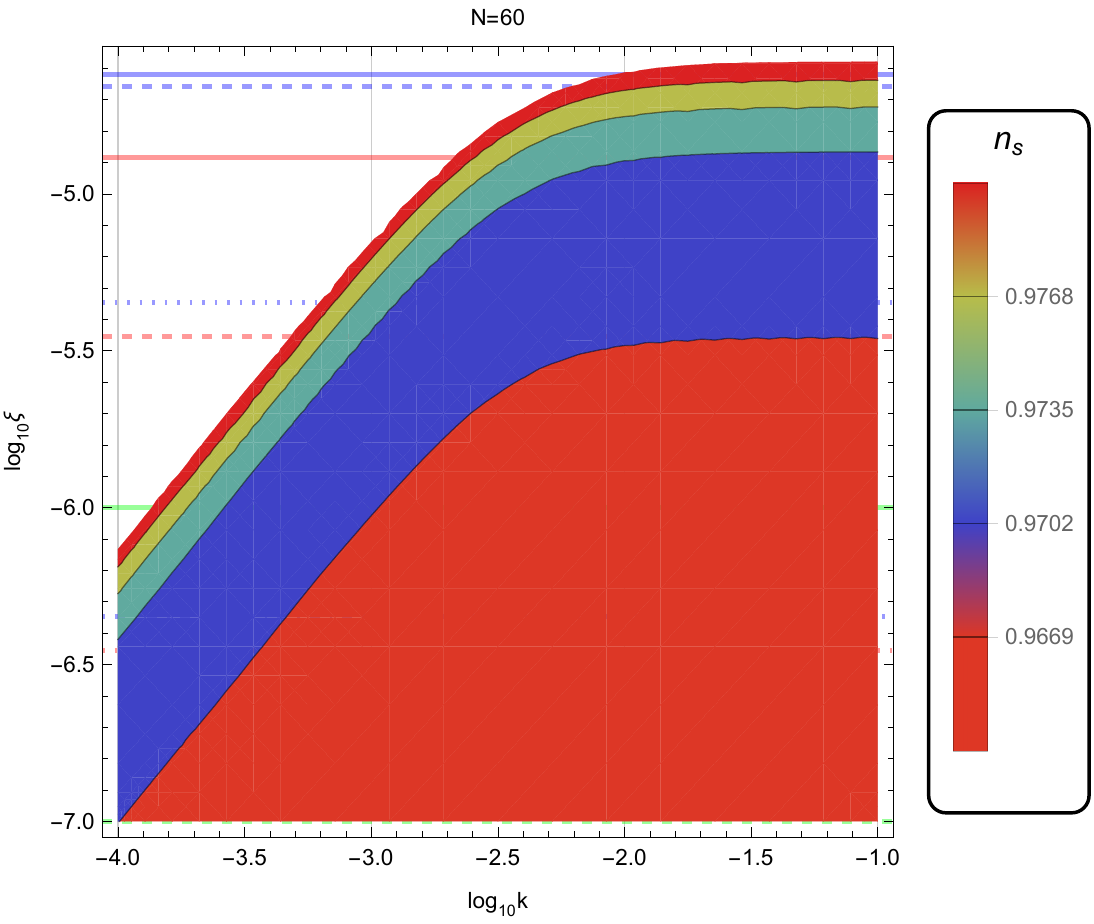}\label{sl_ke-b5}}
    \subfigure[caption]
    {\includegraphics[scale=0.7]{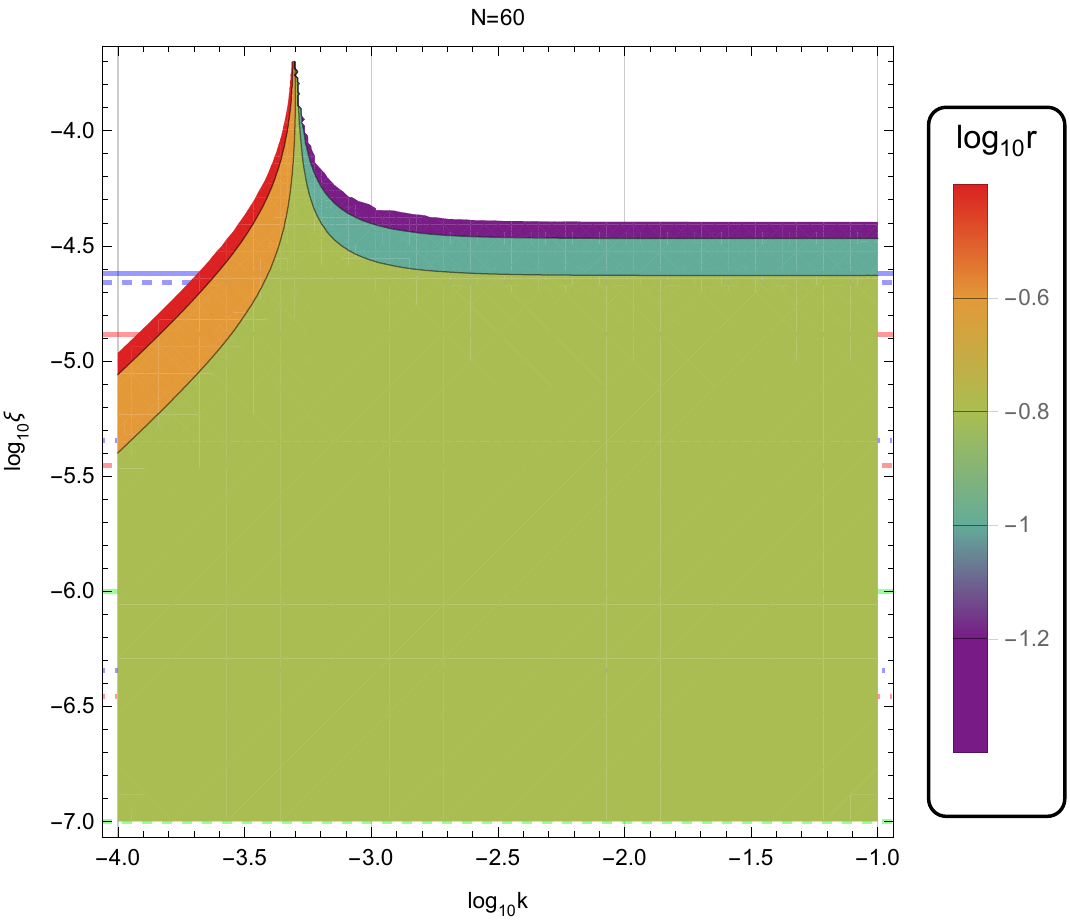}\label{sl_ke-b5}}
    \\
    \subfigure[caption]
    {\includegraphics[scale=0.65]{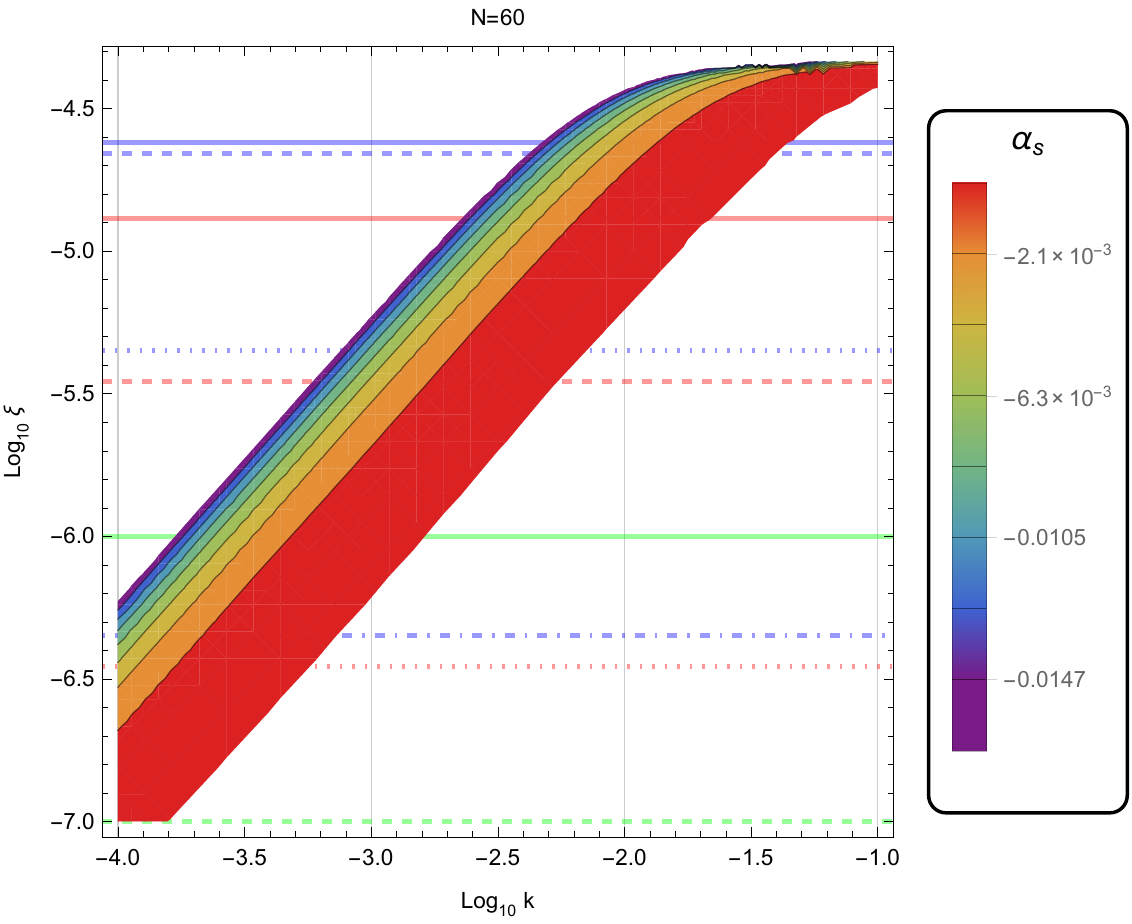}\label{sl_ke-b3}}
    \caption{The $k$ vs. $\xi$\, plot for observable parameters; (\ref{eq:ns}), (\ref{eq:alphas}) and (\ref{eq:ttsrr}). We set $\lambda=0.5\times10^{-12}$, $\phi_i=16M_p$, $a_i=1$ and $M_p^2=1$. For the background grid lines, we set model parameters as follows; the blue solid line is for $\xi=2.4\times10^{-5}$, the blue dashed line is for $\xi=2.2\times10^{-5}$, the red solid line is for $\xi=1.3\times10^{-5}$, the blue dotted line is for $\xi=4.5\times10^{-6}$, the red dashed line is for $\xi=3.5\times10^{-6}$, the green solid line is for $\xi=10^{-6}$, the blue dot-dashed line is for $\xi=4.5\times10^{-7}$, the red dotted line is for $\xi=3.5\times10^{-7}$ and the green dashed line is for $\xi=10^{-7}$, respectively from the top. }
    \label{fig3}
\end{figure}

In Fig.~\ref{fig3}, we plot $k$ vs. $\xi$  for the observable quantities, (\ref{eq:ns}), (\ref{eq:alphas}) and (\ref{eq:ttsrr}). We also plot the grid lines in the background of Fig.~\ref{fig3} where we use exactly same numerical values for $\xi$ that used for Fig.~\ref{fig1}. For the background grid lines, we set model parameters as follows; the blue solid line is for $\xi=2.4\times10^{-5}$, the blue dashed line is for $\xi=2.2\times10^{-5}$, the red solid line is for $\xi=1.3\times10^{-5}$, the blue dotted line is for $\xi=4.5\times10^{-6}$, the red dashed line is for $\xi=3.5\times10^{-6}$, the green solid line is for $\xi=10^{-6}$, the blue dot-dashed line is for $\xi=4.5\times10^{-7}$, the red dotted line is for $\xi=3.5\times10^{-7}$ and the green dashed line is for $\xi=10^{-7}$, respectively from the top. The plot range of each observable parameters chosen such a way that the result to be consistent with observational data~\cite{Ade:2015lrj}. Observational data favor the range of $0.96\leq n_s\leq 0.98$, $10^{-3}\leq r<0.3$ and $-0.017 \leq\alpha_s\leq-0.0002$. As $k$ value decreases in Fig.~\ref{fig3}, the model parameter $\xi$ decreases as well to be consistent with the observational data.

\begin{figure}[H]
    \centering
    {\includegraphics[scale=0.85]{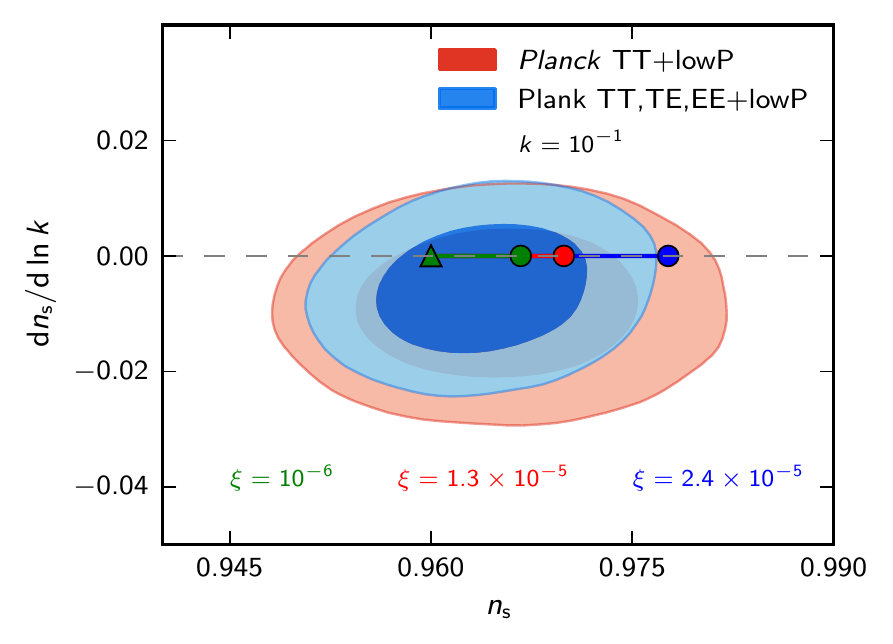}\label{sl_ke-a2}}
    {\includegraphics[scale=0.85]{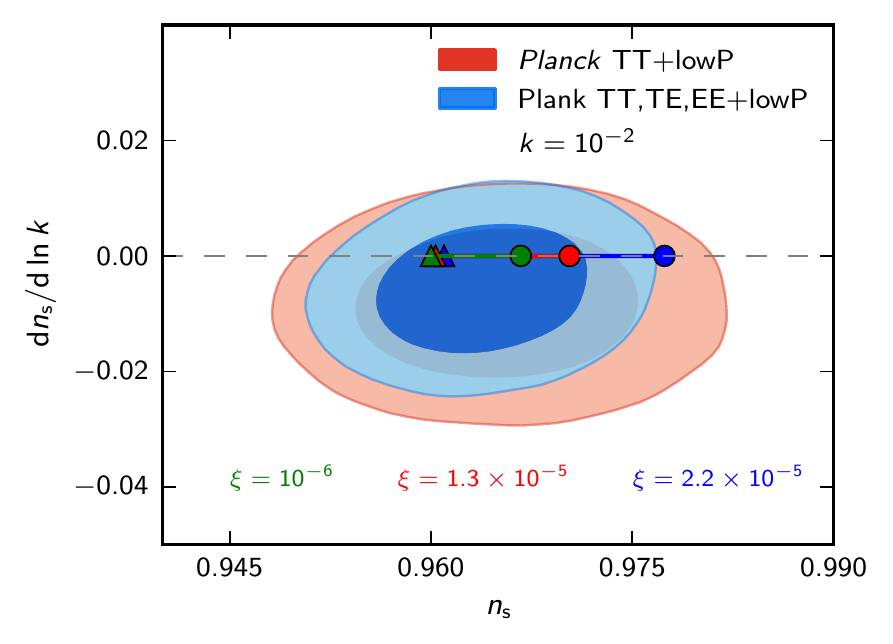}\label{sl_ke-a3}}
    \\
    {\includegraphics[scale=0.85]{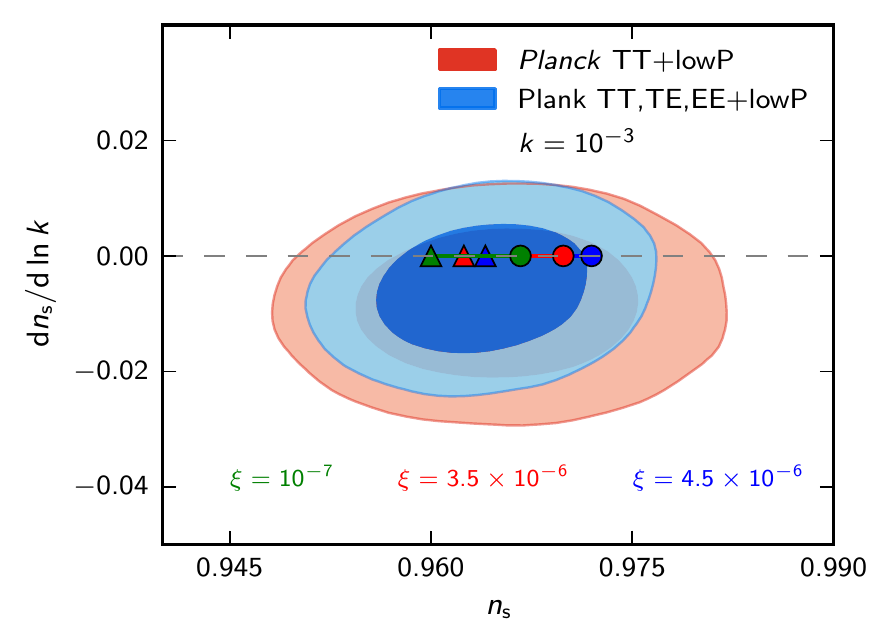}\label{sl_ke-a4}}
    {\includegraphics[scale=0.85]{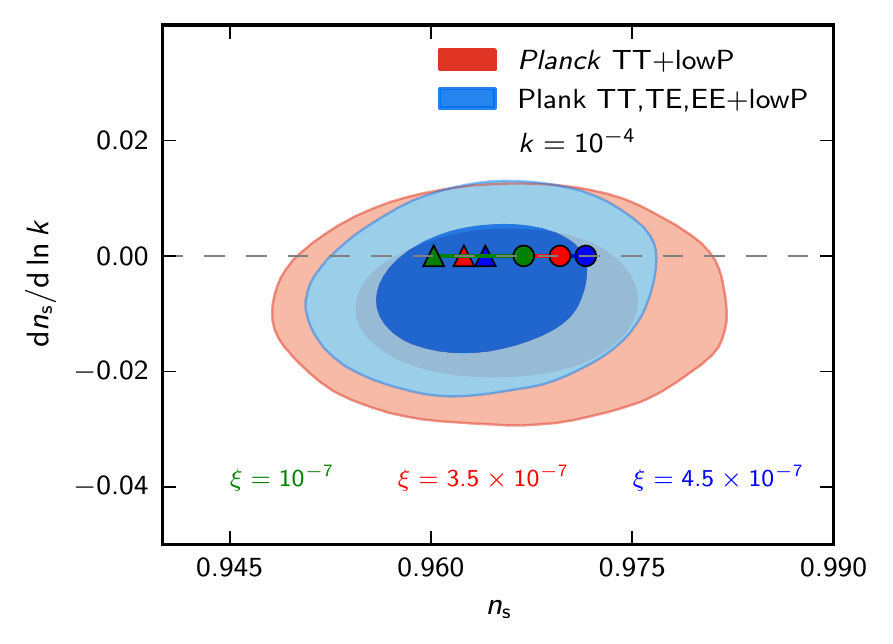}\label{sl_ke-a5}}
    \caption{The $n_s$ vs. $\alpha_s$ plot for $\lambda=0.5\times10^{-12}$, $\phi_i=16$, $a_i=1$ and $M_p^2=1$ where the circle represents $N=60$ while $N=50$ the triangle indicates $N=50$.}
    \label{fig2}
\end{figure}
In Fig.~\ref{fig2}, the plot of $n_s$ vs. $\alpha_s$ for the same numerical values of $\xi$ and $k$ that used in Fig.~\ref{fig1}. Previously in Fig.~\ref{fig1}, it is shown that our result situate outside of $2\sigma$ contour for $N=50$. For the running of scalar spectral indices, the model expectation is now consistent with the data and even situates well inside $2\sigma$ contour for all chosen $\xi$ values.

\section{Conclusion and Discussion} \label{conclusion}
We have studied slow-roll inflation with the nonlinear sigma fields which have
$SO(3)$ symmetry. Motivated from the compactification in  higher dimension theory,
we use the  linearly spatial coordinate dependent solution ansatz
 for the nonlinear sigma fields. There are two interesting features for this model.
First, the $\xi$ term  dominant phase due to the nonlinear sigma fields
in the Friedmann equation is followed by a standard  inflation  phase.
The equation of state parameter for the nonlinear sigma fields gives
$-1/3$, so inflation does not occur when the nonlinear sigma fields are dominant.
This pre-inflation phase would provide or constrain on the initial condition
to the inflation \cite{Kouwn:2014aia}. Second, there exists the
generic cutoff scale in the power spectrum. This cutoff scale originates from
the fact that the comoving horizon is given by $ aH  \sim \kappa \xi/\sqrt{2}$
in the nonlinear sigma field dominant phase.
This implies the mode satisfying $k < k_{min} \sim \kappa \xi/\sqrt{2}$
cannot contribute to the observed power spectrum \cite{koh2015}.

 In light of Planck 2015 observational data,
 we constrain on our model  with $V(\phi) \sim \phi^n$, especially
 for $n =2$. We use the iteration method to obtain the background solution for $\phi$
 and $a$ and then express the slow-roll parameters in terms of $e$-folding number.
For $n_s-r$ plot, our model shows different results depending on $\xi$ as well as $k$
value. For example,  if $10^{-6} < \xi < 2.4 \times 10^{-5}$ and $k=10^{-1}$ for $N=60$
the model predictions reside inside of $2\sigma$ contour.
 If $\xi \leq 10^{-6} \sim 10^{-7}$, the model predictions
give the same results with  usual single field inflation ($\xi =0$).
We have found in Fig. \ref{fig2} that as $k$ value decreases, $\xi$ decreases
as well to be consistent with the observational data

Though we restrict our analysis to the $\phi^2$ potential in this paper,
it would be interesting to consider the different type of potential and
to find the $\xi$ and the potential parameter range to be consistent with observation.
It would also be   interesting to consider inflation with the nonlinear sigma fields
that is coupled with the inflaton.

\section*{Acknowledgements}
S.K. was supported by the Basic Science Research Program through the NRF funded by the Ministry
of Education (No. NRF-2014R1A1A2059080).
B.H.L was supported by the National Research Foundation of Korea(NRF) grant funded by
the Korea government(MSIP) No. 2014R1A2A1A01002306 (ERND).


\end{document}